\documentclass[conference]{IEEEtran}
\IEEEoverridecommandlockouts

\newcommand{\bfh}{{\boldsymbol h}}

\newcommand{\diag}{\mathrm{diag}}
\newcommand{\tr}{\mathrm{tr}}

\newcommand{\bfn}{{\boldsymbol n}}

\newcommand{\bfp}{{\boldsymbol p}}

\newcommand{\bfr}{{\boldsymbol r}}

\newcommand{\bfA}{{\boldsymbol A}} 
\newcommand{\bfB}{{\boldsymbol B}}
\newcommand{\bfC}{{\boldsymbol C}}
\newcommand{\bfD}{{\boldsymbol D}}

\newcommand{\bfQ}{{\boldsymbol Q}}
\newcommand{\bfR}{{\boldsymbol R}}

\newcommand{\bfT}{{\boldsymbol T}}

\newcommand{\bfW}{{\boldsymbol W}}

\newcommand{\rirs}{\boldsymbol{R}_{\text{RIS}}} 
\newcommand{\rtx}{\boldsymbol{R}_{\text{Tx}}}
\newcommand{\Cd}{\boldsymbol{C}_{\mathrm{d}}}
\newcommand{\Cr}{\boldsymbol{C}_{\bfr}}
\newcommand{\Phimat}{{\boldsymbol{\it{\Phi}}}} 
\newcommand{\phivec}{\pmb{\phi}} 
\newcommand{\gammalb}{\gamma^{\mathrm{lb}}}
\newcommand{\channelT}{{\sqrt{\beta}}\rirs^{1/2} {\bfW} \rtx^{1/2, \mathrm{H}}} 

\newcommand{\nm}{\mathrm{n}}
\newcommand{\dm}{\mathrm{d}}
\newcommand{\Tm}{\mathrm{T}}

\newcommand{\nc}{\mathcal{N_\mathbb{C}}}

\newcommand{\Hm}{{\mathrm{H}}}
\newcommand{\Rirs}{\bfR_{\text{RIS}}}
\newcommand{\Rtx}{\bfR_{\text{Tx}}}

\usepackage{cite}
\usepackage{amsmath,amssymb,amsfonts}
\usepackage{algorithmic}
\usepackage{lipsum}
\usepackage{graphicx}
\usepackage{subcaption}
\usepackage{hyperref}
\usepackage{textcomp}
\usepackage{xcolor}
\def\BibTeX{{\rm B\kern-.05em{\sc i\kern-.025em b}\kern-.08em T\kern-.1667em\lower.7ex\hbox{E}\kern-.125emX}}

\usepackage{tikz}
\usepackage{pgfplots}
\pgfplotsset{compat=newest}
\pgfplotsset{plot coordinates/math parser=false}
\newlength\figH  
\newlength\figW  
\usetikzlibrary{quotes,arrows.meta}
\usepackage{physics}
\usepackage{amssymb}
\usepackage{amsmath}
\usepackage{amsthm}

\usepackage[english]{babel}

\usepackage{graphicx}
\usepackage{wrapfig}
\usepackage{caption}
\usepackage{array}
\usepackage{multirow}
\usepackage{arydshln}
\usepackage{color}
\usepackage{xcolor}
\usepackage{tcolorbox}
\usepackage[framemethod=tikz]{mdframed}

\newcommand{\tikznode}[2]{%
	\ifmmode%
	\tikz[remember picture,baseline=(#1.base),inner sep=0pt] \node (#1) {$#2$};
	\else
	\tikz[remember picture,baseline=(#1.base),inner sep=0pt] \node (#1) {#2};%
	\fi}

\definecolor{darkgreen}{rgb}{0,.39,0}

\definecolor{darkred}{rgb}{0.4,0,0}

\definecolor{dnared}{rgb}{1,.27,0} 
\definecolor{dnablue}{rgb}{0,1,1}
\definecolor{dnapurple}{rgb}{0.82,0.13,0.56}
\definecolor{dnagreen}{rgb}{0,1,0}

\usepackage{tikz}

\usetikzlibrary{%
	arrows,%
	automata,%
	backgrounds,%
	calc,%
	patterns,%
	plotmarks,%
	positioning,%
	shapes.geometric,%
	shapes,%
	snakes,%
	decorations.pathmorphing,%
	decorations.shapes,%
	graphs,%
	chains,%
	arrows.meta, %
}

\usepackage{verbatim}
\usepackage{tabularx}
\usepackage{lipsum}
\usepackage{stackengine}
\usepackage{nicefrac}

\usepackage{svg}

\definecolor{bostonuniversityred}{rgb}{0.8, 0.0, 0.0}

\usepackage{xcolor}

\usepackage{mathdots}

\begin{document}

\title{Design of a Single-User RIS-Aided MISO System Based on Statistical Channel Knowledge}

\author{\IEEEauthorblockN{Sadaf Syed, Dominik Semmler, Donia Ben Amor, Michael Joham, Wolfgang Utschick }
\IEEEauthorblockA{{School of Computation, Information and Technology, Technical University of Munich, Germany}\\
Emails: \{sadaf.syed, dominik.semmler, donia.ben-amor, joham, utschick\}@tum.de}
\thanks{\copyright This work has been submitted to the IEEE for possible publication. Copyright may be transferred without notice, after which this version may no longer be accessible.
}
}

\maketitle

\begin{abstract}
Reconfigurable intelligent surface~(RIS) is considered a prospective technology for beyond fifth-generation (5G) networks to improve the spectral and energy efficiency at a low cost. Prior works on the RIS mainly rely on perfect channel state information~(CSI), which imposes a huge computational complexity. This work considers a single-user RIS-assisted communication system, where the second-order statistical knowledge of the channels is exploited to reduce the training overhead. We present algorithms that do not require estimation of the CSI and reconfiguration of the RIS in every channel coherence interval, which constitutes one of the most critical practical issues in an RIS-aided system.  
\end{abstract}

\section{Introduction}
Massive multiple-input multiple-output (MIMO) systems can meet the ever-increasing demand of high throughput and low energy consumption in current wireless communication systems. 
However, equipping the base station~(BS) with a large number of antennas may lead to high circuit energy consumption, including very high hardware costs. Recently, reconfigurable intelligent surface (RIS) has emerged as a promising low-cost solution to enhance the spectral efficiency in a wireless communication system \cite{wu2021intelligent}. 
Specifically, an RIS is a passive array composed of a large number of reconfigurable reflecting elements. Each passive element of the RIS is able to introduce a phase shift to the incident signal in a controlled manner, thereby boosting the received power for the desired user or creating a destructive interference for the non-intended users. Additionally, the passive elements of the RIS do not require any transmit radio frequency (RF) chain, and hence, their energy and hardware costs are much lower as compared to that of the traditional active antennas at the BS. Thus, they can be scaled much more easily than the antennas at the BS.

Most of the existing algorithms for RIS rely on the assumption of perfect channel state information~(CSI), e.g.,~\cite{wu2021intelligent,zhang2020capacity, wu2019intelligent}. 
However, owing to the passive structure of the RIS as well as its massive number of reflecting elements, the acquisition of perfect CSI for the RIS-associated links is formidable. Moreover, these algorithms demand the joint optimisation of the phase shifts and the transmit filters to be performed in every channel coherence interval, which is computationally very expensive. This issue is being recently studied in the literature \cite{hu2020statistical, twotime, statCSI, dang2020joint}, where the key idea is to exploit the statistical knowledge of the channels to design the phase shifts of the RIS. Since the structure of the channels varies slowly, the covariance matrices remain constant for many channel coherence intervals, and hence, it is possible to obtain accurate information of the second-order statistics of the channels through long-term observation. The phase shifts and the filters which are designed based on the covariance matrices do not need to be updated regularly, i.e., there is no need to estimate the channels and perform the joint optimisation in every channel coherence interval. This significantly reduces the channel training overhead and the design complexity of the RIS-assisted systems. The algorithms proposed in \cite{twotime} and~\cite{statCSI} consider the statistical knowledge of the channels for the phase-shift optimisation, however, they consider a hybrid online/offline approach. The phase shifts of the RIS are designed considering the long-term statistics of the channels during the offline step, whereas the filters are designed considering the perfect knowledge of the instantaneous CSI in the online step, thereby, requiring the channel to be estimated perfectly in every channel coherence interval again. 
\par In this work, we present two low-complexity algorithms for a single-user RIS-aided multiple-input single-output~(MISO) system, which are only based on the statistical knowledge of the channels. These algorithms employ the lower bound of the user's rate as the figure of merit, which is based on the worst-case noise bound~\cite{medard}. We consider a more realistic setup, where the covariance matrices of the channels are known perfectly, however, the accurate knowledge of the instantaneous CSI is not available. The bilinear precoders~\cite{amor2020bilinear} are used as the transmit filters, for which a closed-form solution of the optimal filters can be obtained for the single-user case. As such, the filters and the phase shifts can be designed jointly. The algorithm in~\cite{hu2020statistical} is also based on the statistical knowledge of the channels for a single-user MISO system, however, it is based on the assumption that the RIS is deployed at a favourable location and a line-of-sight~(LOS) channel exists to both the BS and the user. The phase shift optimisation in \cite{hu2020statistical} is only dependent on the LOS components, which are assumed to be perfectly known. In this work, we consider a general zero-mean channel model with perfectly known covariance matrices. We compare our algorithms to the one presented in~\cite{dang2020joint}, which assumes a similar zero-mean channel model for a multi-antenna single-user system. The algorithm in~\cite{dang2020joint} maximises the upper bound of the user's rate, which is computed using the Jensen's inequality and it is based on the alternating optimisation~(AO) approach, where the filters and the phase shifts are optimised alternatingly in each subproblem. Such an AO method offers a good performance but it has convergence and complexity issues (discussed in~\cite{guo2020weighted}).   
 

\section{System Model}
This paper investigates the downlink (DL) of an RIS-aided single-user MISO communication system. The system consists of one BS equipped with $M$ antennas, serving one single-antenna user, and one RIS having $N$ passive reflecting elements. The phase-shift matrix of the RIS is defined by a diagonal matrix $\Phimat$ = $\diag$($\phi_1,\cdots,\phi_N$), where $\phi_1, \cdots, \phi_N$ are the phase shift coefficients of the $N$ elements of the RIS with $|\phi_n| = 1 \:\forall \:n$, and $\phivec = [\phi_1, \cdots, \phi_N]^{\Tm}$ denotes the corresponding phase-shift vector. The direct channel from the BS to the user is denoted by $\bfh_{\dm}\in \mathbb{C}^{M\times 1}$, and it is assumed to be circularly symmetric, complex Gaussian distributed with zero mean and covariance matrix $\bfC_{\dm}$, i.e., $\bfh_{\dm}\sim\nc({\bf{0}}, \bfC_{\dm})$. The channel from the RIS to the user is denoted by $\bfr\in \mathbb{C}^{N \times 1}$, which has a zero mean and the covariance matrix $\bfC_{\bfr}$. The channel from the BS to the RIS is denoted by $\bfT \in \mathbb{C}^{N\times M}$, and it is assumed to follow the Kronecker channel model, given by 
\begin{align}
\label{eqn1}
{\bfT} = \channelT. 
\end{align}
The entries of $\bfW\in \mathbb{C}^{N \times M}$ are independent and identically distributed with unit variance and zero mean. $\bfR_{\text{RIS}}$ and $\bfR_{\text{Tx}}$ denote the channel correlation matrices on the side of the RIS and the BS respectively, and $\beta \geq 0$ represents the scaling factor such that $\tr(\Rtx) = \tr\left(\Rirs\right)$ is satisfied. The effective channel of the RIS-assisted system is given by
\begin{align}
\label{eqn6}
\bfh^{\Hm} = \bfh_{\dm}^{\Hm} +  \bfr^{\Hm}\Phimat^{\Hm}\bfT 
\end{align}
which has zero mean and its covariance matrix is given by $\bfC$. It is assumed that the BS has only access to a noisy channel observation $\pmb{\psi}$, but not the actual CSI. The observation $\pmb{\psi}$ is the Least-Squares~(LS) estimate of the channel, which is obtained by correlating the received signal with the pilot sequences during the training phase, and is given by 
\begin{align}
\label{eqn37}
\pmb{\psi} = \bfh \:+\: \bfn
\end{align}
where $\bfn\sim\nc({\bf{0}}, \bfC_{\nm})$ denotes the noise in the channel observation and $\bfC_{\nm}$ is the noise covariance matrix. 

 The transmit filter at the BS is designed such that it only depends on the channel statistics and the noisy observation. To this end, the bilinear precoder \cite{amor2020bilinear} is used as the transmit filter in this work. The bilinear precoder ($\bfp$) is designed such that it linearly depends on the observation $\pmb{\psi}$, i.e., $\bfp~=~\bfA\pmb{\psi}$, with $\bfp~\in \mathbb{C}^{M\times 1}$ and $\bfA \in \mathbb{C}^{M\times M}$ being a deterministic~transformation matrix, which needs to be designed such that the user's rate is maximised. The signal received by the user reads as: $y = \bfh^{\Hm}\bfp\:s + v$,
where $s\sim\nc(0, 1)$ denotes the data symbol and $v \sim\nc(0, \sigma^2)$ is the noise at the user's side. 

Because of the imperfect CSI, we cannot compute the closed-form expression of the actual rate of the user. Instead of that, a lower bound on the user's rate based on the worst-case error, which is extensively used in the massive MIMO literature is employed here \cite{medard}. The lower bound of the user's rate is given by $\log_{2}(1 + \gamma^{\mathrm{lb}})$, where $\gamma^{\mathrm{lb}}$ is the lower bound of the actual signal-to-noise-ratio~(SNR), expressed as
\begin{align}
\label{eqn39}
\gamma^{\mathrm{lb}}  &= \frac{|\mathop{{}\mathbb{E}}[\bfh^{\Hm}\bfp]|^2}{\mathop{{}\mathbb{E}}[|{{\bfh^{\Hm}}{\bfp}} -\mathop{{}\mathbb{E}}[\bfh^{\Hm}\bfp]|^2] + \sigma^2}. 
\end{align}
Evaluating the terms in (\ref{eqn39}) yields (cf. \cite{amor2020bilinear}, \cite{neumann2018bilinear})
\begin{align}
\label{eqn41}
{\gamma}^{\mathrm{lb}} = \frac{|\tr({\bfA\bfC})|^2}{\tr({\bfA \bfQ \bfA^{\Hm}\bfC}) + \sigma^2}
\end{align}
where $\bfQ = \mathop{{}\mathbb{E}}[\pmb{\psi}{{\pmb{\psi}}^{\Hm}}] = \bfC + \bfC_{\nm}$ is the covariance matrix of the LS estimate of the channel. Note that the above closed-form expression of the lower bound is obtained with the Gaussian assumption of $\bfh$, which is indeed true for a large $N$ \cite{wang2022massive}. The matrices $\bfC$ and $\bfQ$ implicitly depend on the phase-shift vector $\phivec$ (shown in the next section). The objective is to maximise the user's rate w.r.t. $\phivec$ and the transformation matrix $\bfA$ of the bilinear precoder. Since the logarithm is a monotonically non-decreasing function, maximising the rate is equivalent to maximising the SNR. Hence, the rate maximisation can be equivalently written as
\begin{align}
&\!\max\limits_{\bfA, \phivec}      &\qquad& \gamma^{\mathrm{lb}} \nonumber\\
&\text{s.t.} &      &  \mathop{{}\mathbb{E}}[||{\bfp}||^2] = \tr(\bfA \bfQ \bfA^{H}) \leq P \label{eqn8b} \tag{P1}\\
&                  &      & |\phi_n| = 1 \:\forall \:n = 1, \cdots, N.  \nonumber
\end{align}

\section{Joint Optimisation Problem Formulation}
Problem \eqref{eqn8b} is non-convex, and hence, it is difficult to obtain a closed-form solution. We next propose theorems to simplify \eqref{eqn8b} such that the filter and the phase shifts can be optimised jointly. 
\subsection{Simplification of the Objective Function}
{\it{\bf{Theorem 1}}}: For a fixed phase-shift vector $\phivec$ of the RIS, the optimal transformation matrix $\bfA \in \mathbb{C}^{M\times M}$ maximising the SNR expression in (\ref{eqn41}) and satisfying the DL power constraint $\mathop{{}\mathbb{E}}[||{\bfp}||^2] \leq P$ for a positive definite matrix $\bfC$ is given by
\begin{align}
\label{eqn42}
\bfA_{\mathrm{opt}} = \eta\:\bfQ^{-1}, \quad \text{where} \: \eta =  \sqrt{\dfrac{P}{\mathrm{tr(\bfQ^{-1})}}}.
\end{align}
\begin{proof}
The SNR expression in (\ref{eqn41}) is a positive real quantity, hence, Wirtinger derivatives are used to find $\bfA$ maximising $\gamma^{\mathrm{lb}}$, which yields $\bfA_{\mathrm{opt}} = \eta\:\bfQ^{-1}$.
Further, $\eta$ can be found from the DL power constraint $\mathrm{tr(\bfA \bfQ \bfA^{H})} = P$.
\end{proof}

Now replacing $\bfA$ in (\ref{eqn41}) with the optimal transformation matrix, the lower bound of the SNR expression becomes
\begin{align}
\label{eqn45}
\gamma^{\mathrm{lb}} = \dfrac{\eta^2\tr^2\left(\bfQ^{-1}\bfC \right)}{\eta^2\:\tr\left(\bfQ^{-1}\bfC\right) + \sigma^2}.
\end{align}

{\it{\bf{Theorem 2}}}: The lower bound of the SNR given in \eqref{eqn45} increases monotonically with $\tr(\bfQ^{-1}\bfC)$ for a spatially white noise covariance matrix $\bfC_{\nm} = \zeta^{2}{\bf{I}}_{M}$ with $\zeta^{2} > 0$. 
\begin{proof}
Please refer to Appendix A.
\end{proof}

Since $\gammalb$ is monotonically increasing with tr($\bfQ^{-1} \bfC$), it is sufficient to maximise tr($\bfQ^{-1} \bfC$). Rewriting $\bfQ^{-1} \bfC$ as ${\bf{I}}_{M}~-~\bfQ^{-1}\bfC_{{\nm}}$ along with the assumption of $\bfC_{\nm}$ to be spatially white, i.e., $\bfC_{\nm} = \zeta^{2}{\bf{I}}_{M}$, \eqref{eqn8b} can be simplified to
\begin{subequations}
\begin{alignat}{2}
&\!\min\limits_{\phivec}      &\qquad& \tr(\bfQ^{-1}) \:\: \text{s.t.}  \:\:\:\: |\phi_n| = 1 \:\forall \:n = 1, \cdots, N. \label{P2} \tag{P2}  
\end{alignat}
\end{subequations}
To solve \eqref{P2}, we first need to express $\bfQ$ as a function of $\phivec$ explicitly. 
\subsection{Computation of the Channel Covariance Matrix}
The channel covariance matrix of the effective channel can be computed as
\begin{align}
         \bfC = \mathop{{}\mathbb{E}}\left[\bfh\bfh^{\Hm}\right] & = \mathop{{}\mathbb{E}} \left[(\bfh_{\dm} + \bfT^{\Hm}\Phimat\bfr)(\bfh_{\dm} + \bfT^{\Hm}\Phimat\bfr)^{\Hm}\right] \\
        &\stackrel{(a)}{=} {\bfC_{\dm}} + \mathop{{}\mathbb{E}} \left[\bfT^{\Hm}\Phimat \bfr \bfr^{\Hm}\Phimat^{\Hm}\bfT\right]
\end{align}
where ($a$) follows from the fact that the random variables $\bfh_{\dm}$, $\bfT$ and $\bfr$ are mutually independent with zero mean, and $\bfh_{\dm}\sim~\nc({\bf{0}}, \bfC_{\dm})$. \\Inserting the expression of $\bfT$ from \eqref{eqn1}, the covariance matrix of the effective channel can be written as
 \begin{align}
      \bfC & \stackrel{(b)}{=} \Cd + \beta \mathop{{}\mathbb{E}} \left[\rtx^{1/2} \bfW^{\mathrm{H}}\rirs^{1/2, \mathrm{H}}\Phimat \Cr \Phimat^{\mathrm{H}}\rirs^{1/2} \bfW \rtx^{1/2, \mathrm{H}}\right] \nonumber
 \end{align}
where ($b$) follows from the fact that $\bfr$ and $\bfW$ are independent random variables, and $\bfr\sim\nc(\bf{0}, \bfC_{\bfr})$. Since the entries of $\bfW$ are i.i.d. with zero mean and unit variance, and $\Phimat = \diag(\phivec)$, the above expression can be simplified as
 \begin{align}
  \label{equation49}
    \bfC & = \Cd + \beta {\mathrm{tr}}(\rirs {\Phimat} \Cr {\Phimat}^{\mathrm{H}})\rtx  \\
   & = \Cd + \beta {\mathrm{tr}}\Big(\rirs (\Cr \odot \phivec \phivec^{\mathrm{H}})\Big)\rtx 
 \end{align} 
where $\odot$ denotes the Hadamard product. Using Lemma 1 of Appendix~B, the above expression can be rewritten as  
\begin{align}
 \bfC & = \Cd + \beta \phivec^{\Hm}\left(\rirs \odot \Cr^{\mathrm{T}}\right)\phivec \rtx. 
 \end{align}
 Thus, the covariance matrix of the LS estimate is given by
 \begin{align}
  \bfQ & = \Cd + \beta \phivec^{\Hm}\left(\rirs \odot \Cr^{\mathrm{T}}\right)\phivec \rtx + \bfC_{{\nm}}. \label{eqn14}
 \end{align}
 \section{Low-Complexity Algorithms Depending on the Channel Statistics}
 In this section, we propose two low-complexity algorithms to solve \eqref{P2}.
 \subsection{Algorithm 1: Projected Gradient Descent Method}
 The minimisation problem in \eqref{P2} can be solved by the iterative projected gradient descent method. The gradient of $\mathrm{tr}(\bfQ^{-1})$ w.r.t. $\phivec^*$ is given by [see \eqref{eqn14}]
\begin{align}
\label{eqn56}
\pdv{\phivec^*}({\mathrm{tr}(\bfQ^{-1} )}) = -\beta \mathrm{tr}(\bfQ^{-1} \rtx {\bfQ^{-1}})(\rirs \odot \Cr^{\mathrm{T}})\phivec.     
\end{align}
\par The expression of the gradient in (\ref{eqn56}) depends on $\bfQ^{-1}$, which depends on $\phivec$. This means that the computation of each gradient step would require the update of the $\bfQ$ matrix and henceforth, the computation of the inverse. This can become computationally very expensive if the size of the matrix $\bfQ$ is large, e.g., as in the case of massive MIMO systems. However, this problem can be easily averted by exploiting the structure of the gradient. The matrix $\bfQ^{-1}$ only appears in the term $\mathrm{tr}(\bfQ^{-1} \rtx {\bfQ^{-1}})$. It can be easily observed that the term $\beta \mathrm{tr}(\bfQ^{-1} \rtx {\bfQ^{-1}})$ is a real non-negative quantity which can be included in the step size optimisation, and thus, we do not have to update the $\bfQ$ matrix after each gradient step. This significantly reduces the computational complexity. The phase shift update rule can hence be summarised as
\begin{align} \label{eqn57}
\phivec \leftarrow \phivec + \kappa\:(\mathbf \rirs \odot \bfC_{\bfr}^{T})\phivec
\end{align}
where $\kappa$ is the optimal step size, which can be computed by the Armijo rule \cite{armijo}. The new phase-shift vector obtained after every gradient step in \eqref{eqn57} should be normalised to satisfy the unit modulus constraints of \eqref{P2}. 
\subsection{Algorithm 2: Element-Wise Optimisation}
The objective function in \eqref{P2} can be reformulated such that it only depends on the $n$-th element of $\phivec$, i.e., $\phi_{n}$, and the remaining $N - 1$ elements are kept fixed in a particular iteration step. To this end, the final expression of $\bfQ$ from \eqref{eqn14} can be rearranged such that it explicitly depends on $\phi_{n}$.
\begin{align}  
 \bfQ & = \Cd + \beta \bigg(\sum_{i = 1}^{N} \sum_{j = 1}^{N}\phi_{i}^{*}\phi_{j}\big[\rirs \odot \Cr^{\mathrm{T}}\big]_{i,j}\bigg)\rtx + \bfC_{{\nm}}. \nonumber
 \end{align}
Rearranging the above equation, we get
\begin{align}
   \bfQ  = \bfD + \phi_{n}\bfB_{n} + {\phi_{n}^{*}}\bfB_{n}^{\mathrm{H}} 
\end{align}
where the matrices $\bfD$ and $\bfB_{n}$ are independent of $\phi_{n}$, and are given by 
\begin{align}
    \begin{split}
    \bfD & = \Cd + \beta\sum_{\substack{i = 1 \\ i \neq n}}^{N} \sum_{\substack{j = 1 \\ j \neq n}}^{N}\phi_{i}^{*}\phi_{j}\big[\rirs \big]_{i,j} \big[\Cr \big]_{j,i}\rtx \\ & + \beta \big[\rirs \big]_{n,n} \big[\Cr \big]_{n,n}\rtx + \bfC_{{\nm}} \\
    \bfB_{n} & = \beta \sum_{\substack{i = 1 \\ i \neq n}}^{N}\phi_{i}^{*}\big[\rirs \big]_{i,n} \big[\Cr \big]_{n,i}\rtx.
    \end{split}    
\end{align}
The optimisation problem in \eqref{P2} can now be reduced to
\begin{subequations}
\begin{alignat}{2}
&\!\min\limits_{\phi_{n}}      &\qquad& \tr(\bfQ^{-1}) \:\: \text{s.t.}  \:\:\:\: |\phi_n| = 1. \label{P3} \tag{P3}  
\end{alignat}
\end{subequations}
The Lagrangian function for the above problem reads as
\begin{align}
\label{eqn73}
{\mathcal{L}} = \mathrm{tr}(\bfQ^{-1}) + \mu(\phi_{n}\phi_{n}^{*} - 1)
\end{align}
where $\mu \in \mathbb{R}$ is the dual variable corresponding to the unit modulus constraint in \eqref{P3}. Solving $\dfrac{\partial {\mathcal{L}}}{\partial \phi_{n}^{*}} \stackrel{!}{=} 0$, we get a closed-form update rule of $\phi_n$ as follows 
\begin{align}
\label{eqn79}
\phi_{n} \leftarrow \dfrac{\mathrm{tr}(\bar{\bfQ}^{-1}{\bfB^{\mathrm{H}}_{n}}\bar{\bfQ}^{-1})}{|{\mathrm{tr}(\bar{\bfQ}^{-1}{\bfB^{\mathrm{H}}_{n}}\bar{\bfQ}^{-1})}|}
\end{align}
where $\bar{\bfQ}$ denotes the value of $\bfQ$ from the previous iteration. In this approach, we do not need to find the optimal step size as in Algorithm~1. However, after each update step, the matrices $\bfQ$ and $\bfB_{n}$ need to be updated, which would be computationally expensive for large $M$, as in the case of massive MIMO systems.
\section{Results}
In this section, numerical results are provided to validate the effectiveness of the proposed algorithms. The system consists of one BS equipped with $M = 4$ antennas, serving one single-antenna user. The RIS is equipped with $N = 40$ reflecting elements. The setup is illustrated in Fig. \ref{setup1}. The user is placed at a distance $D\,\text{m}$ from the BS. Each of the channels is generated according to its distribution as defined in Section~II. The covariance matrix of each channel is generated according to the urban micro channel model described in the 3GPP technical report~\cite{etsi5138}. 
\begin{figure}	
    \centering
    \begin{tikzpicture}[node distance=2cm]
       \fill[violet] (-4,-2) -- node[below, black] {BS (0 m, 0 m)}  (-3,-2) -- (-3.5,-1) -- cycle ;
       \draw (2.5,-1.8) circle (0.2cm) node[label=below:$\text{User ({\it{D}} m, 0 m)}$]{}; 
       \draw[step=0.2cm,black,thin, xshift=0cm,yshift=0cm] (-1,-0.6) grid (0.8,0.6) node[midway, above,yshift=-1.15cm]{$\text{RIS ({50} m, 10 m)}$};
        \draw[-{Stealth[length=2.5mm, width=1.5mm]},dashed] (-3.5,-1.8) -- (2.3,-1.8) node[midway, below]{$\bfh_{\dm}$};
        \draw[-{Stealth[length=2.5mm, width=1.5mm]},dashed] (-3.5,-1.8) -- (-1,0.10) node[midway,sloped, above]{$\bfT$};
        \draw[-{Stealth[length=2.5mm, width=1.5mm]},dashed] (0.9,0.0) -- (2.5,-1.6) node[midway,sloped, above]{$\bfr$};
         \draw[-{Stealth[length=3mm, width=2mm]}] (-3.5,-1.8) -- (-3.5,0.7);
         \draw[-{Stealth[length=3mm, width=2mm]}] (2.7,-1.8) -- (3.5,-1.8);
    \end{tikzpicture}
    \caption{Simulation Setup}
    \label{setup1}
\end{figure}
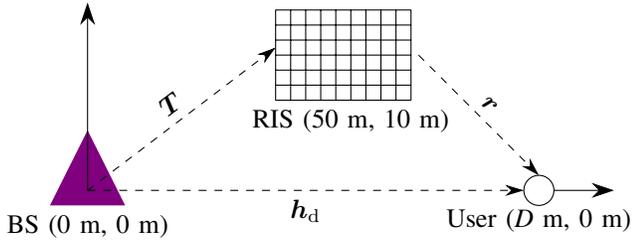
\begin{figure}		
%
%
\definecolor{mycolor1}{rgb}{1.00000,0.00000,1.00000}%
\begin{tikzpicture}

\begin{axis}[%
width=0.85\figW,
height=\figH,
at={(0\figW,0\figH)},
scale only axis,
xmin=0,
xmax=20,
xlabel style={font=\color{white!15!black}},
xlabel={$\bf{Iteration\:No.}$},
ymin=0.1,
ymax=0.28,
ylabel style={font=\color{white!15!black}},
ylabel={$\bf{\tr(\bfQ^{-1})}$},
axis background/.style={fill=white},
xmajorgrids,
ymajorgrids,
legend style={legend cell align=left, align=left, draw=white!15!black, row sep=0cm, font=\tiny}
]
\addplot [color=blue, line width=1.0pt, mark=diamond, mark options={solid, blue}]
  table[row sep=crcr]{%
1	0.261603198833967\\
2	0.250908367983493\\
3	0.15602733794925\\
4	0.15302733794925\\
5	0.14998419086553\\
6	0.14698419086553\\
7	0.139682944630442\\
8	0.133482858547925\\
9	0.124082858547925\\
10	0.119082858547925\\
11	0.118982858547925\\
12	0.118912858547925\\
13	0.118882858547925\\
14	0.118882858547925\\
15	0.118882858547925\\
16	0.118882858547925\\
17	0.118882858547925\\
18	0.118882858547925\\
19	0.118882858547925\\
20	0.118882858547925\\
};
\addlegendentry{$\bf{Algorithm\:1}$}

\addplot [color=red, line width=1.0pt, mark=o, mark options={solid, red}]
  table[row sep=crcr]{%
1	0.261603198833967\\
2	0.131492105763481\\
3	0.119533203671638\\
4	0.116434028345943\\
5	0.115755070368883\\
6	0.115516123598353\\
7	0.115404179438207\\
8	0.115344474959102\\
9	0.115310343373816\\
10	0.11528985805067\\
11	0.115277110103056\\
12	0.115268988480979\\
13	0.115263753228772\\
14	0.115260369013452\\
15	0.115258188019428\\
16	0.115256791738763\\
17	0.115255905573358\\
18	0.115255348652873\\
19	0.115255002258497\\
20	0.115254789072019\\
};
\addlegendentry{$\bf{Algorithm\:2}$}

\end{axis}

\begin{axis}[%
width=1.227\figW,
height=1.227\figH,
at={(-0.16\figW,-0.135\figH)},
scale only axis,
xmin=0,
xmax=1,
ymin=0,
ymax=1,
axis line style={draw=none},
ticks=none,
axis x line*=bottom,
axis y line*=left
]
\end{axis}
\end{tikzpicture}%
		\caption{Convergence Plot for $D$ = $20\,\text{m}$}
		\label{convergence}
\end{figure}
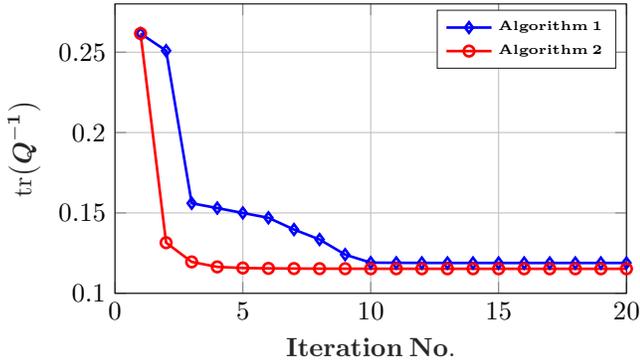
For $D = 20\,\text{m}$, the convergence plot of the two proposed algorithms is shown in Fig.~(\ref{convergence}). The convergence analysis reveals that both algorithms converge in a few iterations. The element-wise optimisation algorithm converges in less than 4 iterations, and the gradient descent based algorithm requires slightly more iterations to converge. It is also observed that the low complexity gradient descent algorithm converges to a similar value as the element-wise optimisation method.      

\begin{figure}
%
%
\definecolor{mycolor1}{rgb}{1.00000,1.00000,0.00000}%
\definecolor{mycolor2}{rgb}{1.00000,0.00000,1.00000}%
\definecolor{mycolor3}{rgb}{0.00000,1.00000,1.00000}%
\begin{tikzpicture}

\begin{axis}[%
width=0.9\figW,
height=\figH,
at={(0\figW,0\figH)},
scale only axis,
xmin=-20,
xmax=40,
xlabel style={font=\color{white!15!black}},
xlabel={$\bf{Transmit\:Power\:in\:dB}$},
ymin=0,
ymax=16,
ylabel style={font=\color{white!15!black}},
ylabel={$\bf{User's\:Rate\:[bpcu]}$},
axis background/.style={fill=white},
xmajorgrids,
ymajorgrids,
legend style={at={(0.01,0.3)}, anchor=south west, legend cell align=left, align=left, draw=white!15!black, row sep=-0.05cm, font=\tiny}
]

\addplot [color=black, dashed, line width=1.0pt, mark=o, mark options={solid, black}]
  table[row sep=crcr]{%
-20	0.0856328101505123\\
-10	0.663134558869106\\
0	2.65825420862607\\
10	5.78579556515837\\
20	9.04407056491944\\
30	12.3172479267124\\
40	15.7270658836213\\
};
\addlegendentry{$\bf{SDR\:(Perfect\:CSI)\:[3]}$}

\addplot [color=black, dashdotted, line width=1.0pt, mark=triangle, mark options={solid, rotate=90, black}]
  table[row sep=crcr]{%
-20	0.0199901263594281\\
-10	0.191010318538022\\
0	1.10613037195198\\
10	3.59055065607716\\
20	6.57154145799079\\
30	9.84980249708097\\
40	13.1662752126901\\
};
\addlegendentry{$\bf{SDR\:(Imperfect\:CSI)\:[3]}$}

\addplot [color=green, dashdotted, line width=1.0pt, mark=o, mark options={solid, green}]
  table[row sep=crcr]{%
-20	0.0076090218856916\\
-10	0.0624779925157209\\
0	0.49560472805375\\
10	2.01925549793462\\
20	4.5502395378743\\
30	7.53421236511627\\
40	10.8811736584046\\
};
\addlegendentry{$\bf{AO\:Algorithm\:[7]}$}

\addplot [color=red, line width=1.0pt, mark=o, mark options={solid, red}]
  table[row sep=crcr]{%
-20	0.0237416698777577\\
-10	0.174609305038857\\
0	0.942964010825341\\
10	3.14910053705563\\
20	5.88740850268363\\
30	9.04305469829078\\
40	12.556205018454\\
};
\addlegendentry{$\bf{Algorithm\:2}$}

\addplot [color=mycolor3, dashed, line width=1.0pt]
  table[row sep=crcr]{%
-20	0.001485323377815\\
-10	0.012894606416807\\
0	0.119161278435432\\
10	0.821270900461883\\
20	2.71908261130914\\
30	5.55532442664717\\
40	8.82659676904828\\
};
\addlegendentry{$\bf{Random\:Phase\:Shift}$}

\addplot [color=yellow, dashed, line width=1.0pt]
  table[row sep=crcr]{%
-20	0.001485323377815\\
-10	0.012894606416807\\
0	0.074161278435432\\
10	0.571270900461883\\
20	2.21908261130914\\
30	5.05532442664717\\
40	8.32659676904828\\
};
\addlegendentry{$\bf{No\:RIS\:[9]}$}

\addplot [color=blue, line width=1.0pt, mark=diamond, mark options={solid, blue}]
  table[row sep=crcr]{%
-20	0.0192191789497158\\
-10	0.14237156206731\\
0	0.731237139568615\\
10	2.92233654369326\\
20	5.64565107467853\\
30	8.78384319454364\\
40	12.2980683895856\\
};
\addlegendentry{$\bf{Algorithm\:1}$}

\addplot [color=red, dashed, line width=1.0pt, mark=diamond, mark options={solid, red}]
  table[row sep=crcr]{%
-20	0.0264445413197507\\
-10	0.224661799317446\\
0	1.28624338053411\\
10	3.98378366268583\\
20	7.05109097636356\\
30	10.3399885476025\\
40	13.7603688983208\\
};
\addlegendentry{$\bf{TTS\:with\:Algorithm\:2}$}

\addplot [color=mycolor2, dashed, line width=1.0pt, mark=o, mark options={solid, mycolor2}]
  table[row sep=crcr]{%
-20	0.0156550530730704\\
-10	0.146954016332464\\
0	0.994931437356182\\
10	3.40483434274334\\
20	6.60175057488962\\
30	9.77482163513496\\
40	13.0106716362337\\
};
\addlegendentry{$\bf{TTS\:[5]}$}

\end{axis}

\begin{axis}[%
width=1.227\figW,
height=1.227\figH,
at={(-0.16\figW,-0.135\figH)},
scale only axis,
xmin=0,
xmax=1,
ymin=0,
ymax=1,
axis line style={draw=none},
ticks=none,
axis x line*=bottom,
axis y line*=left
]
\end{axis}
\end{tikzpicture}%
   \caption{User's Rate vs Transmit Power $P$ in dB}
   \label{plot2}
\end{figure}
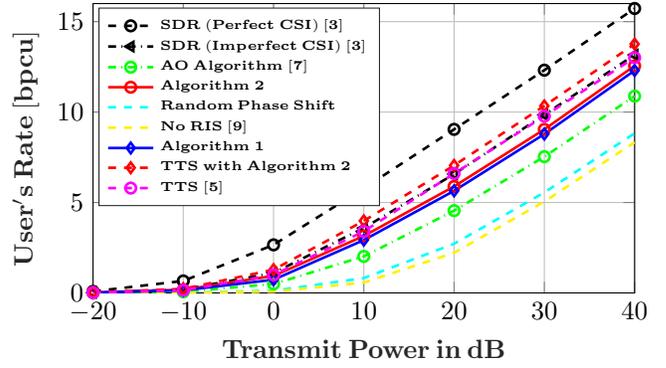

The user's rate is taken as the performance metric in Fig.~(\ref{plot2}), which is computed with the different algorithms and compared over the transmit power levels $P$. The average rate of the user is given by $\mathop{{}\mathbb{E}}\big[\log_2(1 + |\bfh^{\Hm}\bfp|^2/\sigma^2)\big]$, where $\sigma^2$ is set to 1. The estimation noise covariance matrix $\bfC_{\nm}$ is assumed to be the identity matrix. The rate is averaged over 100 covariance matrices, which are generated by varying the distance $D$ in between $15\,\text{m}$ to $60\,\text{m}$ and the path loss factors of the scatterers randomly. For each of the generated covariance matrices, the user's instantaneous rate is averaged over 1000 channel realisations. The performance of the proposed algorithms is compared with the following baselines: (i) a system without RIS with the bilinear precoders as the transmit filters \cite{amor2020bilinear}, (ii) a system with RIS where the phase shifts are chosen randomly and the bilinear precoders are used as the transmit filters, (iii) the SDR approach of~\cite{wu2019intelligent} for the genie-aided setup of perfectly known CSI, (iv) the SDR approach of~\cite{wu2019intelligent} used for the imperfect CSI setup, (v) the algorithm in \cite{dang2020joint} based on the statistical channel knowledge, and (vi) the two-timescale~(TTS) approach of~\cite{twotime}. 
 Fig.~(\ref{plot2}) compares the user's rate for the different schemes with respect to the transmit power $P$ in dB. 
The topmost curve represents the upper bound of the rate that can be achieved for the considered system setup when the CSI is perfectly known, and the optimisation of filters and phase shifts is performed in every channel coherence interval with the SDR method~\cite{wu2019intelligent}. The SDR algorithm of~\cite{wu2019intelligent} is then employed in an imperfect CSI setup and the user's rate degrades by 9~dB approximately. The simulation results reveal that the two proposed algorithms are very similar in performance. Moreover, their performance gap to the SDR approach for the imperfect CSI scenario is small, despite the fact that these algorithms are computationally much less expensive as the filters and the phase shifts do not need to be optimised in every channel coherence interval. Furthermore, these algorithms based on the maximisation of the lower bound of the user's rate considering the worst-case noise bound \cite{medard} outperform the AO algorithm in~\cite{dang2020joint}, which maximises the upper bound of the rate obtained through Jensen's inequality. Additionally, we extend the algorithms to the TTS approach of~\cite{twotime}. The algorithm in \cite{twotime} employs the stochastic successive convex approximation~(SSCA) method \cite{liu2018online} to compute the optimal phase shifts based on the channel statistics. In the TTS approach, the optimal phase shifts obtained by \eqref{eqn57}, \eqref{eqn79} or the SSCA method~\cite{twotime} are kept fixed in the coherence interval of the covariance matrices and the filters are updated in every channel coherence interval with the matched filter~(MF). It is observed that the TTS approach employing Algorithm~2 outperforms the algorithm in~\cite{twotime} for our system setup, i.e., the performance of the TTS optimisation is boosted by the method underlying Algorithm~2 and it offers the best performance among other approaches involving the statistical channel knowledge in Fig.~(\ref{plot2}).             
  
\section{Conclusion}
In this work, we have presented algorithms for the single-user RIS-aided MISO systems based on the bilinear precoders. The simulation results illustrate that a performance gain can be achieved by optimising the phase shifts of the RIS, even when the actual CSI is not available, by exploiting the second-order statistics. This significantly reduces the training overhead as the channels do not need to be estimated in every channel coherence interval and the phase shifts of the RIS do not need to be updated frequently. The extension of the algorithms for the multi-user setup will be presented in our next work. 
\section{Appendix}
\subsection{Proof of Theorem 2}
 With $\eta =  \sqrt{\dfrac{P}{\mathrm{tr(\bfQ^{-1})}}}$, $\gamma^{\mathrm{lb}}$ can be rewritten as
  \begin{align}
 \label{equation18}
 \gamma^{\mathrm{lb}} =\dfrac{\tr^2(\bfQ^{-1}\bfC)}{\tr(\bfQ^{-1}\bfC) + \sigma^{2}\:\tr(\bfQ^{-1})/P}\:.
  \end{align}
  Assuming $\bfC_{\nm}$ = $\zeta^{2}\:{\bf{I}}_M$, where $\zeta^2 > 0$, the term $\tr(\bfQ^{-1})$ can be written as $\tr(\bfQ^{-1}(\bfQ - \bfC))/\zeta^{2}$, which, in fact, equals to $\Big(M - \tr(\bfQ^{-1}\bfC) \Big)/\zeta^{2}$. Plugging this into~(\ref{equation18}), and replacing the term $\tr(\bfQ^{-1}\bfC)$ by $x$ for the ease of notation, the lower bound of the SNR can be expressed as a function of $x$ by
  \begin{align}
 \label{equation19}
 \gamma^{\mathrm{lb}} = f(x) & = \dfrac{x^2}{{\left(1 - \dfrac{\sigma^2}{P\zeta^2}\right)}\:x + {\dfrac{\sigma^{2}M}{P\zeta^{2}}}}.
  \end{align}
Replacing $\left(1 - \dfrac{\sigma^2}{P\zeta^2}\right)$ by $k_1$ and $\dfrac{\sigma^{2}M}{P\zeta^{2}}$ by $k_2$, we get
\begin{align}
    f(x) = \dfrac{x^{2}}{k_1\:x + k_2}\:\text{and}\:f'(x) = \dfrac{{k}_{1}x^{2} + 2\:k_{2}\:x}{(k_1\:x + k_2)^2}.
\end{align}
It can be easily observed that $x = \tr(\bfQ^{-1}\bfC)$ is always positive because $\bfQ$ and $\bfC$ are positive definite matrices. Hence, we are interested in the sign of the term ${k}_{1}x + 2\:k_{2}$ to determine the sign of $f'(x)$. Also, note that $k_2 > 0$ since $M, \:P,\:\zeta^{2},\:\sigma^{2}>0$.
\par \noindent \underline{Case 1}: $P\zeta^{2} - \sigma^2 \geq 0$, i.e., $k_1 \geq 0$. \\
It is easy to verify that $f'(x) > 0$ for this case. 
 \par \noindent \underline{Case 2}: $P\zeta^{2} - \sigma^2 < 0$, i.e., $k_1<0$. 
 \begin{align}
\label{equation23}
{k}_{1}x + 2\:k_{2} & = \left(1 - \dfrac{\sigma^2}{P\zeta^2}\right)\tr(\bfQ^{-1}\bfC)  +  \dfrac{2\:\sigma^{2}M}{P\zeta^{2}} \nonumber\\
&\stackrel{(a)}= M + \dfrac{\sigma^{2}M}{P\zeta^{2}} - k_1 \tr(\bfQ^{-1}\bfC_{\nm}) > 0
\end{align}
 where ($a$) follows from $\bfC = \bfQ - \bfC_{\nm}$. This shows that $f'(x)~>~0$ holds for this case too. Hence, $f(x)$ is always monotonically increasing in $x$. This proves Theorem~2.
\subsection{Lemma 1}
For any three matrices $\bfA$, $\bfB$ and $\bfC$ of the same dimensions, we have
\begin{align}
\label{equation36}
\mathrm{tr}\Big(\bfA (\bfB \odot \bfC)\Big) =  \mathrm{tr}\Big((\bfA \odot \bfB^{\mathrm{T}}) \bfC\Big).
\end{align}
\begin{proof}
\begin{align*}
  \mathrm{tr}\Big(\bfA (\bfB \odot \bfC)\Big) & = \sum_{i}\big[\bfA (\bfB \odot \bfC)\big]_{i,i} \\
 & = \sum_{i}\bigg(\sum_{k}\big[\bfA \big]_{i,k}\:\big[\bfB \big]_{k,i}\: \big[\bfC \big]_{k,i}\bigg)
\end{align*}
\begin{align*}
 \mathrm{tr}\Big((\bfA \odot \bfB^{\mathrm{T}}) \bfC\Big) & = \sum_{i}\big[(\bfA \odot \bfB^{\mathrm{T}}) \bfC \big]_{i,i}\\
& = \sum_{i}\bigg(\sum_{k} \big[\bfA \big]_{i,k}\:\big[\bfB \big]_{k,i}\: \big[\bfC \big]_{k,i}\bigg)
\end{align*}
Hence, L.H.S. = R.H.S., and this proves Lemma 1.
\end{proof} 
\bibliographystyle{IEEEtran}
\bibliography{bibliography}
\end{document}